\documentclass[sigplan,10pt,nonacm]{acmart}

\makeatletter
\def\@ACM@checkaffil{% Only warnings
    \if@ACM@instpresent\else
    \ClassWarningNoLine{\@classname}{No institution present for an affiliation}%
    \fi
    \if@ACM@citypresent\else
    \ClassWarningNoLine{\@classname}{No city present for an affiliation}%
    \fi
    \if@ACM@countrypresent\else
        \ClassWarningNoLine{\@classname}{No country present for an affiliation}%
    \fi
}
\makeatother

\renewcommand\footnotetextcopyrightpermission[1]{}
\AtBeginDocument{%
  \providecommand\BibTeX{{%
    \normalfont B\kern-0.5em{\scshape i\kern-0.25em b}\kern-0.8em\TeX}}}
\usepackage{underscore}
\usepackage{caption}
\usepackage{subcaption}
\usepackage[subtle]{savetrees}

\usepackage{listings}
\usepackage[inline]{enumitem}
\setlist{noitemsep,nolistsep}

\setlist[itemize]{leftmargin=*}
\setlist[enumerate]{leftmargin=*}

\begin{document}

%%
%% The "title" command has an optional parameter,
%% allowing the author to define a "short title" to be used in page headers.
\title{The Cost of Garbage Collection \\ for State Machine Replication}

%%
%% The "author" command and its associated commands are used to define
%% the authors and their affiliations.
%% Of note is the shared affiliation of the first two authors, and the
%% "authornote" and "authornotemark" commands
%% used to denote shared contribution to the research.
\author{Zhiying Liang}
\affiliation{%
  \institution{The Pennsylvania State University}
  % \city{University Park}
  % \country{United States}
}
\email{zvl5490@psu.edu}

\author{Vahab Jabrayilov}
\authornote{Work was done as an intern at The Pennsylvania State University}
% \orcid{0000-0002-1825-0097}
\affiliation{%
  \institution{Columbia University}
  % \city{New York}
  % \country{United States}
}
\email{vj2267@columbia.edu}

\author{Aleksey Charapko}
% \orcid{0000-0001-5109-3700}
\affiliation{%
  \institution{University of New Hampshire}
  % \city{Durham}
  % \country{United States}
}
\email{Aleksey.Charapko@unh.edu}

\author{Abutalib Aghayev}
\affiliation{%
  \institution{The Pennsylvania State University}
  % \city{University Park}
  % \country{United States}
}
\email{agayev@psu.edu}

%%
%% By default, the full list of authors will be used in the page
%% headers. Often, this list is too long, and will overlap
%% other information printed in the page headers. This command allows
%% the author to define a more concise list
%% of authors' names for this purpose.
\renewcommand{\shortauthors}{Trovato and Tobin, et al.}

\renewcommand{\sectionautorefname}{\S} \renewcommand{\subsectionautorefname}{\S}
\renewcommand{\subsubsectionautorefname}{\S}

%%
%% The abstract is a short summary of the work to be presented in the
%% article.
\begin{abstract}
State Machine Replication (SMR) protocols form the backbone of many distributed systems. Enterprises and startups increasingly build their distributed systems on the cloud due to its many advantages, such as scalability and cost-effectiveness. One of the first technical questions companies face when building a system on the cloud is which programming language to use. Among many factors that go into this decision is whether to use a language with garbage collection (GC), such as Java or Go, or a language with manual memory management, such as C++ or Rust. Today, companies predominantly prefer languages with GC, like Go, Kotlin, or even Python, due to ease of development; however, there is no free lunch: GC costs resources (memory and CPU) and performance (long tail latencies due to GC pauses). While there have been anecdotal reports of reduced cloud cost and improved tail latencies when switching from a language with GC to a language with manual memory management, so far, there has not been a systematic study of the GC overhead of running an SMR-based cloud system.

This paper studies the overhead of running an SMR-based cloud system written in a language with GC. To this end, we design from scratch a canonical SMR system---a MultiPaxos-based replicated in-memory key-value store---and we implement it in C++, Java, Rust, and Go. We compare the performance and resource usage of these implementations when running on the cloud under different workloads and resource constraints and report our results. Our findings have implications for the design of cloud systems.
\end{abstract}

%%
%% The code below is generated by the tool at http://dl.acm.org/ccs.cfm.
%% Please copy and paste the code instead of the example below.
%%
% \begin{CCSXML}
% <ccs2012>
%  <concept>
%   <concept_id>10010520.10010553.10010562</concept_id>
%   <concept_desc>Computer systems organization~Embedded systems</concept_desc>
%   <concept_significance>500</concept_significance>
%  </concept>
%  <concept>
%   <concept_id>10010520.10010575.10010755</concept_id>
%   <concept_desc>Computer systems organization~Redundancy</concept_desc>
%   <concept_significance>300</concept_significance>
%  </concept>
%  <concept>
%   <concept_id>10010520.10010553.10010554</concept_id>
%   <concept_desc>Computer systems organization~Robotics</concept_desc>
%   <concept_significance>100</concept_significance>
%  </concept>
%  <concept>
%   <concept_id>10003033.10003083.10003095</concept_id>
%   <concept_desc>Networks~Network reliability</concept_desc>
%   <concept_significance>100</concept_significance>
%  </concept>
% </ccs2012>
% \end{CCSXML}

% \ccsdesc[500]{Computer systems organization~Embedded systems}
% \ccsdesc[300]{Computer systems organization~Redundancy}
% \ccsdesc{Computer systems organization~Robotics}
% \ccsdesc[100]{Networks~Network reliability}

%%
%% Keywords. The author(s) should pick words that accurately describe
%% the work being presented. Separate the keywords with commas.
%\keywords{datasets, neural networks, gaze detection, text tagging}

%%
%% This command processes the author and affiliation and title
%% information and builds the first part of the formatted document.
\settopmatter{printfolios=true}
\maketitle
\pagestyle{plain}

\section{Introduction}

%The cloud's unprecedented value proposition---an immediately available infrastructure at the needed scale---streamlines innovation, which drives its rapid adoption among enterprises and startups.
State machine replication (SMR) has an unprecedented role in modern distributed and cloud computing. SMR algorithms~\cite{raft,paxos-made-live,paxos-made-complex,epaxos,vr} underpin most critical systems and services, ranging from configuration management~\cite{zookeeper,etcd,zeus,delos,million-tiny-dbs} for the cloud infrastructure to massive data-intensive systems that handle petabytes of data and millions of requests per second~\cite{spanner,cockroachdb,yugabytedb,mongodb,dynamodb,paxosstore}. With SMR protocols often on the critical path of data-intensive systems, they need to fulfill a no-compromise role of delivering high throughput with low latency, high reliability, and high efficiency. 

The latter point on efficiency, however, can often be neglected in smaller systems and startups, especially in light of cutting upfront costs of development and maintenance. The cloud's unprecedented value proposition---an immediately available infrastructure at the needed scale---further enables companies to focus on their product and growth first. However, it is increasingly becoming clear that when companies grow on the cloud without considering the resource efficiency of their software infrastructure upfront, the cloud cost soars after some point~\cite{ibm-cloud-cost, dataconomy-cloud-cost} and starts to drive down profit margins, leaving the companies with a dilemma of whether to rewrite or to ``repatriate''---move from the cloud back to on-premise~\cite{a16z-cloud-cost}.

The resource efficiency and performance of data-intensive systems running on the cloud are significantly affected by whether they are implemented in a language with GC or in a language with manual memory management. While the automatic memory management offered by the languages with GC is a boon to programmer productivity~\cite{phipps}, it does not come for free: the language runtime needs to run sophisticated GC algorithms, using a considerable amount of CPU and memory in addition to those used by the application code---translating to a higher monthly cloud bill.

%It is important to emphasize that modern programming languages with GC predate the rise of cloud computing. Thus their designers could not have envisioned the utility nature of the cloud present today and thought about the long-term consequences of opting for GC. Even Go, which is often described as the ``language of the cloud''~\cite{go-cloud, go-cloud2}, was born in 2007~\cite{go-history}, just a year after the announcement of Amazon EC2 in 2006~\cite{aws-history}. Hence, the languages with GC are rooted in the pre-cloud era, in which the extra hardware cost to offset the GC overhead is a one-time expense justified by the increased programmer productivity---not a monthly bill susceptible to inflation~\cite{cloudflation1, cloudflation2}.

Increasingly, growing startups report experiencing GC-caused problems~\cite{discord-go-to-rust,akita,fanatics}, and some re-implement their systems in a language with manual memory management and reap significant benefits. For example, Discord, which is a popular instant messaging platform running on Google Cloud, switched from Go to Rust to avoid long tail latencies~\cite{discord-go-to-rust} and just recently replaced a database written in Java with a clone of the same database written in C++, reducing the size of their database fleet from 177 servers to 72 servers~\cite{discord-scylladb}. While such reports are interesting, they are hard to draw conclusions from because the rewrites may have also introduced improved redesigns, making the comparisons between languages unfair. And although there are many prior works~\cite{gc-vs-manual1, gc-vs-manual2, gc-vs-manual3, lion} that compare and analyze the performance of identical programs written in a language with GC and in a language with manual memory management, (a) these studies are usually restricted to small programs, whose results are hard to extrapolate to the behavior of complex distributed cloud systems with many moving parts, and (b) they are conducted on bare metal, but as we show (\autoref{sec:virt-overhead}), the virtualization overhead on the cloud affects languages differently.

In this paper, we focus on systematically studying the GC costs and efficiency of state machine replication (SMR) systems due to their prevalence and importance for many large systems and services. Specifically, we first design a canonical SMR system---a MultiPaxos-based distributed in-memory key-value store called Replicant---entirely in pseudocode, relying on common data structures and concurrency primitives available in all languages, such as threads, mutexes, and condition variables, and then we port our pseudocode to C++, Java, Rust, and Go. We compare (a) the performance of C++ and Java implementations as the languages with manual memory management and GC, respectively, which natively use operating system (OS) threads, and (b) the performance of Rust and Go implementations as the languages with manual memory management and GC, respectively, which natively use user-mode threads. We run our experiments on AWS using YCSB~\cite{ycsb} as the load generator while varying the available resources. 

We find that GC has a high cost when targeting low tail latency---a requirement for cloud systems~\cite{tail-at-scale}. Specifically, (1) for update-heavy workloads, with ample memory (memory to data ratio = 8), C++ and Rust achieve 1.6$\times$ and 1.4$\times$ higher throughputs than Java and Go, respectively; with limited memory (memory to data ratio = 1.25), C++ achieves 9.6$\times$ higher throughput than Java, and Go is unable to achieve the target tail latency, while Rust's throughput is unchanged from the ample-memory case; (2) for read-heavy workloads, with ample memory, C++ and Rust achieve 1.08$\times$ and 1.33$\times$ higher throughputs than Java and Go, respectively; with limited memory, C++ achieves 1.7$\times$ higher throughput than Java, and Go is unable to achieve the target tail latency, while Rust's throughput is unchanged from the ample-memory case. These results imply significant cloud cost savings in the long run if a continuously growing cloud system is built in a language with manual memory management.

We specifically chose to base Replicant on MultiPaxos because despite being arguably the most impactful distributed protocol powering many proprietary~\cite{chubby,spanner,megastore,dynamodb,borg,paxosstore} and open-source~\cite{ceph,cassandra,foundationdb} systems and influencing new protocols~\cite{raft}, there is no open-source, modular, standalone, reference implementation of MultiPaxos. The academic implementations we found~\cite{epaxos,nopaxos,libpaxos,paxi,franken-paxos, paxos-made-complex} were either too simplistic, glossing over important details, or lacked important features found in production implementations, such as log trimming, which made them unsuitable for our purpose. We concocted MultiPaxos details from multiple papers~\cite{paxos-made-simple,parliament,paxos-made-live,paxos-made-complex} and filled in the ambiguities.

We make the following contributions:
\begin{itemize}
    \item We present the first systematic study of the overhead of running an SMR-based cloud system, Replicant, written in a language with GC on the cloud.% Specifically, we compare and analyze the performance and resource usage of Replicant implementations in C++, Java, Rust, and Go under various workloads with different resource limitations and report our results.
    \item We present the design of Replicant, a clean and modular MultiPaxos-based distributed in-memory key-value store, and its implementations in C++, Java, Rust, and Go. Our design documents and code, consisting of 18{,}926 lines (including unit tests), is at \url{https://github.com/psu-csl/replicated-store}.
    %\item We present a lightweight log compaction mechanism for MultiPaxos that does not require proactive snapshotting.
\end{itemize}

We refrain from making definitive conclusions about the choice of a programming language for building cloud systems because the reality of building a large production system is often complicated, constrained by factors other than just the cloud cost, such as languages lacking certain necessary features, bugs in third-party libraries (as exemplified by our experience in this paper), the availability of engineers knowing the language, and the library ecosystem, to name a few. Instead, we present our results and describe our experience and observations (\autoref{sec:discussion}), providing helpful information for those making decisions.
\section{Background}

An established approach for comparing a particular property of programming languages---be it performance~\cite{gc-vs-manual1, gc-vs-manual2, gc-vs-manual3, lion} or energy efficiency~\cite{couto}---is to create a benchmark, a set of small programs representing typical workloads with \textit{identical} implementations in the languages being compared, and then run the programs and compare them with regards to the studied property. While such works are useful in their own right, it is hard to extrapolate the results from small programs to large and complex cloud systems. Ideally, to study programming language properties as they pertain to cloud systems, one would create a benchmark consisting of typical and realistic cloud systems (as exemplified by DeathStarBench~\cite{deathstar})---e.g., databases, caches, and web servers---and implement them identically in multiple languages and compare them with regards to the studied property under realistic workloads. Creating such a benchmark is a significant undertaking. We take the first step in this direction by creating one such cloud system---a fault-tolerant and linearizable distributed key-value store we call Replicant.

Such key-value stores are at the core of many cloud services, either as standalone systems~\cite{etcd, zookeeper, tikv} or as building blocks for larger systems~\cite{spanner, cockroachdb}. They achieve fault tolerance by running multiple replicas, where each replica acts as a deterministic state machine, and ensuring that all replicas execute the same sequence of deterministic commands---an approach known as State Machine Replication (SMR)~\cite{fbs-tutorial}. SMR is typically implemented with a consensus protocol, which is used for building a \textit{replicated log} among replicas. Specifically, each replica maintains a private copy of the log of commands, and the replicas run a consensus protocol among themselves to agree on the order of commands in the log and subsequently execute them.

Replicant implements SMR using MultiPaxos protocol. We next give a brief description of Single-Decree Paxos, upon which MultiPaxos is based, followed by a description of MultiPaxos. Since MultiPaxos is not well defined in the literature, our description contains some elements of our design.

\subsection{Single-Decree Paxos}
Single-Decree Paxos is a fundamental consensus protocol that ensures a group of peers agree on some \textbf{single value}~\cite{paxos-made-simple}. It operates in three phases and reaches an agreement on a single value among multiple peers as long as the majority of peers are operational.

In the first phase, called the \textit{prepare phase}, a proposer peer \textit{P} that wants to propose a value chooses a ballot number \textit{\textbf{b}} and sends a \textit{prepare} message \textit{\textbf{<prepare, b>}} to all of the peers. If an acceptor peer \textit{Q} receiving the \textit{\textbf{<prepare, b>}} message has previously seen a \textit{\textbf{<prepare, b'>}} message with a ballot number \textit{\textbf{b'}} higher than \textit{\textbf{b}} (i.e.,  $b'>b$), then \textit{Q} rejects \textit{P}'s \textit{prepare} message; otherwise, \textit{Q} accepts \textit{P}'s \textit{prepare} message. Upon acceptance, \textit{Q} needs to reply to \textit{P} with a \textit{promise} message, which serves dual purposes---first, it signals peer ~\textit{P} that \textit{Q} has not seen any higher ballots, and second, it allows \textit{P} to learn of any values that some prior proposer \textit{P'} may have proposed. If \textit{Q} has previously accepted an \textit{accept} message (described in the next paragraph) from another proposer \textit{P'}, which contains a proposed value \textit{\textbf{v}} along with a ballot number \textit{\textbf{c}}, then \textit{Q} responds to \textit{P} with \textit{\textbf{<promise,b,v,c>}}. If \textit{Q} has not accepted any prior \textit{accept} messages, then it can simply reply with \textit{\textbf{<promise,b>}}. If \textit{P}'s \textit{prepare} message is accepted by the majority of peers, then \textit{P} moves to the second phase---the \textit{accept phase}; otherwise, it chooses a higher ballot number and repeats.

Once in the \textit{accept phase}, \textit{P} proposes the value \textit{\textbf{w}} to be chosen; \textit{\textbf{w}} is either a value that \textit{P} learned in the \textit{prepare phase} (if multiple peers respond in the \textit{prepare phase} with different values or ballots then P chooses \textit{\textbf{w}} to be \textit{\textbf{v}} with the highest \textit{\textbf{c}}), or a new value of \textit{P}'s choosing, if \textit{P} did not learn a new value in the \textit{prepare phase}. To propose a value, \textit{P} sends an \textbf{\textit{<accept,w,b>}} message to all peers, containing the value \textit{\textbf{w}} and ballot number  \textit{\textbf{b}} that was accepted by the majority of peers in the \textit{prepare phase}. Each acceptor peer \textit{Q} accepts the \textbf{\textit{<accept,w,b>}} message only if it has not received any messages---\textit{prepare} or \textit{accept}---containing a higher ballot number. If the acceptor \textit{Q} is not aware of a higher ballot number, then it can accept \textit{P}'s \textit{accept} message. If the majority of peers accept \textit{P}'s \textit{accept} message, then the consensus has been reached, and the value \textit{\textbf{w}} has been chosen; otherwise, \textit{P} restarts at the \textit{prepare phase}.

Once \textit{\textbf{w}} is chosen, only \textit{P} is aware of it being chosen. Hence, in the third phase, called the \textit{commit phase} (also sometimes known as the \textit{learn phase}~\cite{mencius}), \textit{P} lets other peers know that \textit{\textbf{w}} was chosen by sending a \textbf{\textit{<commit,w,b>}} message to all peers.

\subsection{MultiPaxos}
\label{sec:multipaxos}
While it is possible to build a replicated log by running Single-Decree Paxos to reach a consensus on the value of the command for each \textit{instance} (i.e., entry) in the replicated log, most implementations~\cite{bolosky, chubby, paxos-made-live, megastore, borg} use an optimization of Single-Decree Paxos, called MultiPaxos, which eliminates most \textit{prepare} messages and achieves consensus in a single phase (the \textit{accept phase}) for most instances in the log.

MultiPaxos is a leader-driven protocol. The leader receives the commands from clients, determines the command order in the replicated log, and replicates these ordered commands to the remaining peers. The protocol uses a \textit{leader election} process to establish one peer as the leader. This process is essentially the \textit{prepare phase} of Single-Decree Paxos modified to work with the replicated log, as opposed to a single value. More specifically, in a leader election, given a set of peers, any of which can propose commands, the first peer that successfully completes the \textit{prepare phase} becomes a \textit{leader} and lets the remaining peers know of its leadership (e.g., by replicating commands or sending heartbeat messages). As a result, the other peers abandon their leadership attempts and become \textit{followers}. 

From then on, the clients send commands to the leader (if a follower receives a command from a client, it redirects the client to the leader). The leader decides the index (i.e., position) \textit{\textbf{i}} of the instance for the command \textit{\textbf{c}} in the log, skips the \textit{prepare} message, and sends the \textbf{\textit{<accept,b,i,c>}} message to peers. Once the leader receives an acceptance from the majority of peers, the consensus has been reached that the command for the instance at index \textit{\textbf{i}} is \textit{\textbf{c}}. The leader then lets the followers know that the command for the instance at index \textit{\textbf{i}} has been chosen by sending them a \textit{commit} message with \textit{\textbf{i}}. Often, the commit message can be piggybacked onto some later \textit{accept} message.

Since clients issue commands concurrently, there are multiple concurrent \textit{accept phases} running at any given time, and due to factors such as network delay, the commands can be committed out of order. However, commands must be executed in the log order to ensure that the state machines in all replicas progress identically. Hence, a peer (either a leader or a follower) can execute command \textit{\textbf{c}} of the instance at index \textit{\textbf{i}}, once all the commands of the instances at indexes up to \textit{\textbf{i}} have been committed and executed. Finally, instances with executed commands must be trimmed from the log to prevent the log's unbounded growth. A standard approach to log trimming is to take the snapshot of the state machine after executing the command at instance \textit{\textbf{j}}, and then trim all the instances up to and including \textit{\textbf{j}}.

MultiPaxos can efficiently reach consensus for a log instance in one round-trip as long as the leader does not fail. If the leader fails, e.g., it temporarily disconnects, then the followers detect this and elect a new leader, which has a ballot number that is higher than the old leader's ballot number; if the old leader then reconnects, it discovers the new leader (either via heartbeats or via receiving a rejection to its messages with new leader's ballot number) and becomes a follower.

As explained above, unlike Single-Decree Paxos, which uses a single ballot number to reach consensus on a single value MultiPaxos uses a single ballot number to reach consensus for multiple log instances. Hence, when a peer runs the \textit{prepare phase} during leader election in MultiPaxos, the other peers respond not with a single value, but with the complete contents of their log. Once a peer completes the \textit{prepare phase} and becomes a leader, it determines the largest index, \textit{\textbf{i\_max}}, in the logs it received, and it immediately starts receiving commands from the clients and running the \textit{accept phase} for the commands starting at index \textit{\textbf{i\_max + 1}}. Concurrently, the leader (1) merges the received logs; if an instance at index \textit{\textbf{i}} exists in multiple logs and it is committed or executed, then the instances at \textit{\textbf{i}} in those logs must match; otherwise, the leader chooses the instance with the highest ballot number to be at \textit{\textbf{i}} in the merged log; and (2) \textit{replays} the merged log by running the \textit{accept phase} for each instance in the merged log. Replaying the merged log is paramount for temporarily disconnected peers to recover the instances they missed while they were disconnected, so that they can catch up the progression and continue executing new commands.

\section{Replicant Architecture}
This section describes the high-level architecture of Replicant and the MultiPaxos module. Our repository contains a detailed design document of these and other modules.

\begin{figure}
    \centering
    \includegraphics[width=\linewidth]{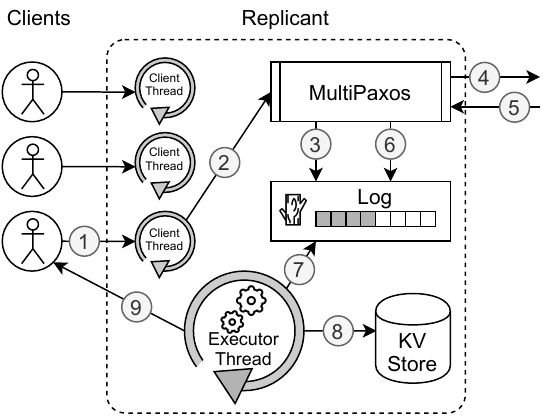}
    \caption{Replicant Design}
    \label{fig:replicant}
\end{figure}

\subsection{Replicant Design}
\label{sec:replicant-design}

\autoref{fig:replicant} shows the high-level design of Replicant, which consists of the MultiPaxos module, the KVStore module, and the Log module. The KVStore module is a wrapper around an in-memory hash table. The Log module is the MultiPaxos protocol log implemented as a thread-safe unbounded producer-consumer queue. Finally, the MultiPaxos module encapsulates the MultiPaxos protocol, which communicates with other MultiPaxos modules embedded in other Replicant peers to reach a consensus on the value of a specific instance in the log.

Replicant creates a new thread for handling each incoming client, assigns a unique id to each client, and maintains a map from the client's id to the client's socket descriptor. \autoref{fig:replicant} shows a typical workflow: (1) a client issues a command to Replicant; (2) a client handler thread passes the command to MultiPaxos; (3) MultiPaxos creates an instance consisting of the command, index of the instance in Log, ballot number, and client-id, and appends the instance to Log; (4) MultiPaxos sends the \textit{accept} message with the instance to all other peers, and (5) once it receives an acceptance from the majority of peers, (6) it commits the instance in Log; finally, when an instance is committed in Log, the Executor Thread (7) extracts the command and the client id from the committed instance, (8) executes the command on KVStore, uses the client id to find the corresponding socket descriptor, and (9) writes the response to the client. While clients can issue commands concurrently, and thus there may be concurrent \textit{accept phases} running in the MultiPaxos module, the execution of commands occurs serially in the Executor Thread to ensure the same sequence of state transitions in all peers.

\subsection{MultiPaxos Design in Replicant}

\begin{figure}
    \centering
    \includegraphics[width=\linewidth]{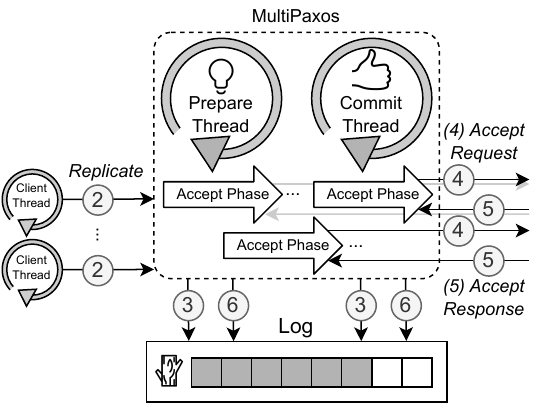}
    \caption{MultiPaxos Design}
    \label{fig:multipaxos}
\end{figure}

We now describe the key aspects of our MultiPaxos design in Replicant; our repository contains a more detailed design document. \autoref{fig:multipaxos} shows the high-level design of the MultiPaxos module in Replicant and its interactions with the Log module; although log is typically considered an integral part of MultiPaxos, we found that having it as a separate module outside of MultiPaxos results in a cleaner design.

\subsubsection{Common Path} The MultiPaxos module has a single public method, \textit{Replicate}, which is called concurrently by the client threads in Replicant as shown in \autoref{fig:multipaxos}; its arguments are the command \textit{\textbf{c}} and the \textit{\textbf{id}} of the client issuing \textit{\textbf{c}}. In the common case, \textit{Replicate} runs the \textit{accept phase} by (3) appending an instance containing \textit{\textbf{<ballot,c,id,state>}} to Log, where \textit{\textbf{state}} is initialized to \textit{in-progress}, (4) sending an \textit{accept} message with the instance to all other peers, (5) receiving acceptance from the majority, and (6) committing the instance by changing its \textit{\textbf{state}} in Log to \textit{committed}. Committing an instance wakes up the Executor Thread, which, if possible, executes the committed instance and responds to the client with the result. If it is impossible to execute the instance due to a gap in Log, e.g., instance 2 has just committed while instance 1 still has not, then when instance 1 eventually gets committed, the Executor Thread will wake up and execute instance 1, followed by instance 2 and respond to the corresponding clients.

\subsubsection{Leader Election} As the \autoref{fig:multipaxos} shows, there are two long-running threads in the MultiPaxos module---the Prepare Thread and the Commit Thread---each named after a MultiPaxos phase that they are executing. Upon startup, the peers run as followers; the Prepare Thread keeps running as long as the peer is a follower, and if the peer stops receiving heartbeats from the leader, the Prepare Thread waits for a random period and runs the \textit{prepare phase}. The random wait increases the likelihood of one of the peers completing the \textit{prepare phase} and establishing its leadership before other peers attempt the \textit{prepare phase} and split the votes, similar to Raft~\cite{raft}. However, unlike Raft's leader election, any peer in MultiPaxos can become a leader, not just the peer with the longest log.

After a peer becomes a leader, the Prepare Thread replays the merged log (\autoref{sec:multipaxos}) and goes to sleep, to be woken up when the peer becomes a follower again, and the Commit Thread wakes up. As long as the peer is a leader, the Commit Thread keeps executing the \textit{commit phase} at a fixed interval (150\,ms by default) by sending \textit{commit} messages to followers, which also act as heartbeats.

Multiple instances can get committed between two \textit{commit} messages. The leader, however, cannot send just the index of the last committed instance to followers because a previous instance may still be in progress, and having the followers commit an instance that is not committed by the leader would violate safety. Thus, the leader should either send the indexes of individual instances that have been committed or alternatively---as we do in our MultiPaxos implementation---it could send just \textit{\textbf{last\_executed}}---the index of the last executed instance, which is guaranteed to have all instances before it committed and executed (and thus it is safe for the followers to start committing and then executing them).

If the leader becomes a follower, as described in \autoref{sec:multipaxos}, then the Commit Thread goes to sleep, to be woken up when the peer becomes a leader again, and the Prepare Thread wakes up again. After several iterations in our design, we found that having separate symmetric threads for each phase resulted in a cleaner and easier-to-follow design.

\section{Implementation and Experiments Setup}

\begin{figure*}
     \begin{subfigure}[b]{0.49\textwidth}
         \includegraphics[width=\linewidth]{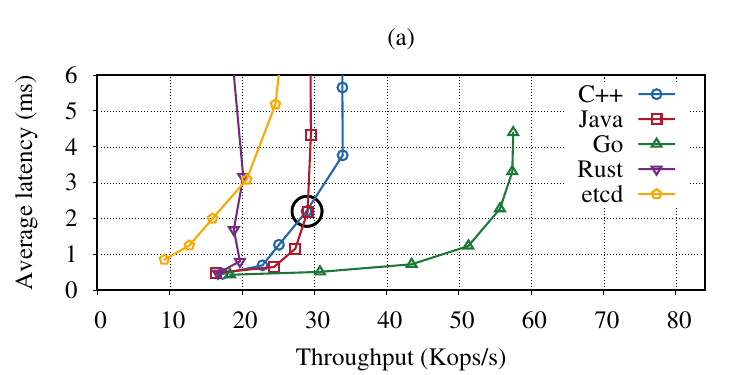}
     \end{subfigure}
     \hfill
     \begin{subfigure}[b]{0.49\textwidth}
         \includegraphics[width=\linewidth]{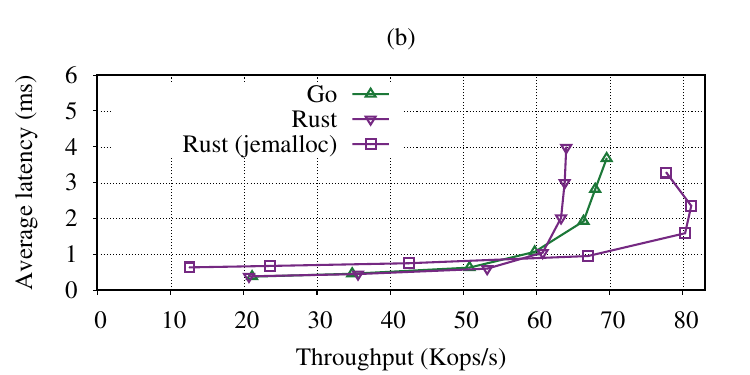}
     \end{subfigure}
        \caption{Throughput vs. average latency of Replicant implementations in different languages, (a) using gRPC for inter-peer communication, and (b) using TCP for inter-peer communication, under YCSB workload A with 2 million entries, which mitigates GC pressure. The data points correspond to 8, 16, 32, 64, 128, 192, and 256 concurrent clients; e.g., the circled data point in (a) corresponds to 64-client results for C++ and Java implementations. Data points with higher than 6\,ms average latency are not shown.}
        \label{fig:grpc-and-tcp-latency}
\end{figure*}

We first designed and implemented Replicant entirely in pseudocode (in multiple iterations over many months), assuming standard concurrency primitives, such as threads, mutexes, and condition variables. Once the design stabilized, the team members concurrently ported the pseudocode into C++, Java, Go, and Rust. The implementations are identical, including in their use of algorithms, data structures, modules, variables, and variable names (sans language-specific conventions, e.g., snake case in Rust and camel case in Java).

\subsection{Threading} Since Go supports only user-mode threads (i.e., goroutines), we initially envisioned using user-mode threads in all languages for a fair comparison---specifically, coroutines in C++~\cite{cppcoroutines}, Tokio tasks in Rust~\cite{tokio}, and Project Loom virtual threads in Java~\cite{javaloom}. However, at the time of our implementation, Project Loom was still in beta, and our microbenchmarks showed lackluster results with Java virtual threads; therefore, we decided to use operating system (OS) threads in C++ and Java and compare the two and use user-mode threads in Rust and Go and compare the two. In retrospect, this decision allowed us to observe the advantage of user-mode threads over OS threads when building systems on the cloud (\autoref{sec:virt-overhead}).

\subsection{Networking} Google's RPC library, gRPC~\cite{grpc}, is the de facto standard for cloud backend servers, and all of our target languages have gRPC implementations (Google maintains gRPC for C++, Java, and Go; for Rust, the Tonic~\cite{tonic} library is a popular and feature-complete gRPC implementation). Therefore, we used gRPC for inter-peer MultiPaxos communication in our first implementation iteration. As described below, we found a scalability bug in Tonic; hence, for a fair comparison, we rewrote our Rust and Go implementations to use plain TCP for inter-peer communication and kept gRPC in our C++ and Java implementations. For client-to-server communication, all implementations use asynchronous TCP.

\subsection{Setup}
\label{sec:setup}
We run all experiments on a 3-server setup with each server running Replicant on an m5.2xlarge instance (8 vCPUs, 32\,GiB RAM) on AWS. We use Clang C++ compiler v14.0.6, OpenJDK v19.0.1, Go v1.19.1, and Rust v1.64.0. As a benchmark, we run YCSB on a separate m5.2xlarge instance with a variable number of client threads. We choose key and value sizes of 23\,B and 500\,B, respectively, based on the recent literature~\cite{twitter1}. We first populate the database with 7.5 million entries (unless otherwise noted) and perform a 20-second warm-up before running each experiment. Our workloads use a Zipfian request distribution. When reporting a throughput, we repeat every experiment multiple times and report the result with error bars; when reporting a  latency CDF, we repeat each experiment multiple times, ensuring they behave similarly, and then report one of the samples.

\subsection{The First Implementation and the TCP Fork}
\label{sec:tcp-fork}
\autoref{fig:grpc-and-tcp-latency}\,{(a)} shows the throughput vs. average latency graph of our first Replicant implementations in four languages using gRPC for inter-peer communication, along with etcd~\cite{etcd}---a production key-value store implemented in Go that uses Raft for its replicated log. Since the purpose of our project is to produce a clean and modular implementation with acceptable performance in different languages and compare their resource usage, we include etcd in the graph not for performance comparison but as a sanity check to show that Replicant's performance is reasonable, and it has a similar asymptotic behavior. Unlike Replicant, which is an in-memory key-value store, etcd is a persistent key-value store; hence, to make the comparison sensible, in this experiment, etcd uses a RAM disk as the persistent store.

The key observation from \autoref{fig:grpc-and-tcp-latency}\,{(a)} is that C++ and Java (for brevity, we refer to a specific Replicant implementation using the language of its implementation), both of which use the OS threads, are closer in their maximum throughputs (29\,Kops/s vs. 33\,Kops/s) compared to Go, which has a higher maximum throughput (57\,Kops/s) and lower average latency at every data point. Go's advantage stems primarily from its use of user-mode threads (we omit our technical analysis to save space, but the theory behind it is well-known~\cite{pressler}). However, we observe that Rust, which also uses user-mode threads, has the worst performance. Our lengthy debugging and profiling of this anomaly revealed that the H2 library~\cite{h2}---the HTTP2 library on which Tonic is based---uses a global lock for managing HTTP2 streams~\cite{h2-lock}, limiting the scalability of Tonic (RPCs in gRPC correspond to HTTP2 streams~\cite{grpc-on-http2}).

We are working with the H2 and Tonic maintainers to remove the global lock in H2. However,  for our further experiments, we forked our code, redesigned and reimplemented the MultiPaxos module in Rust and Go implementations of Replicant to use TCP for inter-peer communication with JSON for message serialization. We omit the detailed description of our redesign (available in our repository), but again, both implementations are identical, including in their use of data structures, algorithms, variables, and variable names.

\autoref{fig:grpc-and-tcp-latency}\,{(b)} shows the throughput vs. latency graph of the Replicant implementations in Rust and Go that use TCP for inter-peer communication. First, we observe that Go now has 19\% higher throughput than its gRPC version (68\,Kops/s vs. 57\,Kops/s), which is expected since gRPC adds significant overhead. More importantly, though, Rust now scales, although Go still outperforms it. Our profiling of the Rust implementation revealed that \texttt{ptmalloc}~\cite{ptmalloc}---the default Linux memory allocator---is the bottleneck. As the figure shows, replacing \texttt{ptmalloc} with \texttt{jemalloc}---a highly optimized allocator developed at Facebook~\cite{jemalloc}---results in significantly higher throughput with 64 or more clients at the cost of lower throughput with 32 or fewer clients. For the remaining experiments, the Rust implementation of Replicant uses \texttt{jemalloc}.% as the allocator

\subsection{Avoiding Coordinated Omission}
\label{sec:co}
Coordinated Omission is a long-known problem in the industry~\cite{co}. In a nutshell, it refers to a load generator's conflation of \textit{service time}, which is the time it takes a server to complete the request, with the \textit{response time}, which is the sum of the \textit{service time} and the \textit{wait time}---the time a request spends in the queue~\cite{response-time}. This conflation happens because many load generators behave as a \textit{closed system}, while they should be behaving as an \textit{open system}~\cite{open-vs-closed} to reflect reality. More specifically, in a closed-system load generator, a new request arrival is triggered by the completion of the previous request; therefore, assuming a server that receives 100 requests per second and completes each of the requests 1-96 in 1\,ms and each of the requests 97-100 in 250\,ms, a closed-system load generator will report the 95th percentile latency as 1\,ms. An open-system load generator, on the other hand, will send the requests at a constant rate, independent of the completion times of the previous requests. Since most requests sent after the first second will get queued at the load generator (due to the last four slow requests in every 100 requests), it will correctly report the 95th percentile latency as 250\,ms or higher.

\begin{figure}
    \centering
    \includegraphics[width=\linewidth]{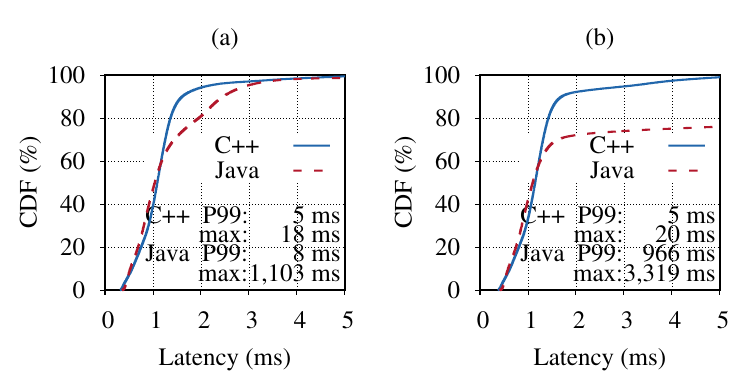}
    \caption{The latency CDFs generated using (a) default and (b) true latency numbers reported by YCSB for Replicant implementations in C++ and Java running the experiment of \autoref{fig:grpc-and-tcp-latency}\,{(a)} with 64 clients and target throughput of 20\,Kops/s, and with server memory limited to 5\,GiB.}
    \label{fig:coordimated-omission}
\end{figure}

Coordinated Omission was first publicized by Gil Tene a decade ago~\cite{co}, and soon afterward, YCSB implemented a fix~\cite{ycsb-co-fix} by adding a new option to measure the true latency---called \textit{intended latency} in YCSB~\cite{ycsb-co-options}. Unfortunately, despite the previous studies on the difference between closed-loop and open-loop clients~\cite{treadmill-co}, this option is not the default and it is almost mentioned in academic works that use YCSB. However, it has a profound effect on the latency reported by YCSB for services that experience hiccups (e.g., GC pauses). We demonstrate this by repeating the experiment of \autoref{fig:grpc-and-tcp-latency}\,{(a)} for C++ and Java using 64 clients with a target throughput of 20\,Kops/s, limiting the memory of servers to 5\,GiB to force GC in Java. In \autoref{fig:coordimated-omission}\,{(a)} we plot the latency CDFs using the default latency numbers reported by YCSB; as the graph shows, the 99th percentile latencies of C++ and Java are close (5\,ms vs. 8\,ms). On the other hand, when we plot the CDFs using the true latency numbers---obtained by passing \texttt{intended} option to the \texttt{measurement.interval} parameter---in \autoref{fig:coordimated-omission}\,{(b}), we observe that Java's 99th percentile latency (966\,ms) is almost 200$\times$ higher than that of C++'s (5\,ms). In the rest of this paper, we report only the true latency numbers.

\subsection{Disabling Nagle's Algorithm}

\begin{figure}
    \centering
    \includegraphics[width=\linewidth]{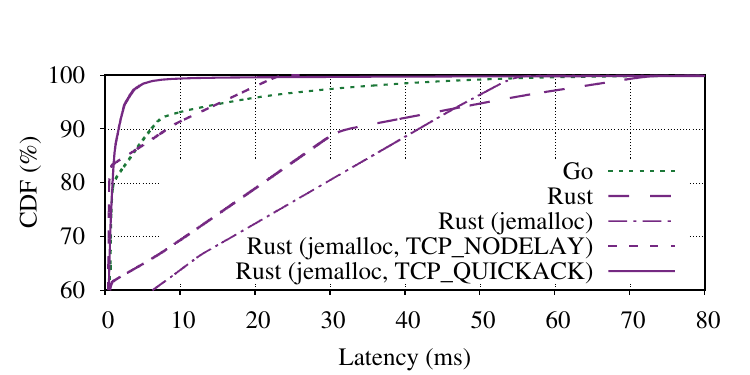}
    \caption{The latency CDFs of Replicant implementations in Go and in Rust (using the stock allocator and jemalloc allocator with Nagle algorithm turned off and with TCP Quick ACK) running the experiment of \autoref{fig:grpc-and-tcp-latency}\,{(b)} with 64 clients and target throughput of 50\,Kops/s.}
    \label{fig:nagle}
\end{figure}

Unlike many languages, Go disables Nagle's algorithm~\cite{nagle} by default via setting TCP\_NODELAY option on a socket~\cite{go-turn-off-nagle}. \autoref{fig:nagle} shows the effect this has on tail latency when running the experiment of \autoref{fig:grpc-and-tcp-latency}\,{(b)} with 64 clients and target throughput of 50\,Kops/s. Go's 99th percentile latency (47\,ms) is lower than that of stock Rust (69\,ms); with \texttt{jemalloc}, Rust gets closer (54\,ms), and with Nagle's algorithm disabled, Rust achieves lower 99th percentile latency (22\,ms) than Go. As an aside, we also show the effect of using TCP Quick ACK (recommended by Nagle as a better alternative to his algorithm~\cite{nagle-hn}), which reduces Rust's 99th percentile latency down to 7\,ms. However, since Go doesn't support TCP Quick ACK, we do not enable it in Rust. In the rest of this paper, we disable Nagle's algorithm in all our implementations for all experiments.

\section{Measuring the GC Overhead}

In this section, we quantify the cost of GC on botha 3-node and 5-node Replicant clusters running on AWS (\autoref{sec:setup}). Since both 3-node and 5-node configurations exhibit similar trends across all experiment settings, we opt to present the results of the 3-node Replicant cluster to avoid redundancy. We first compare C++ and Java as the two languages with manual memory management and GC, respectively, which use OS threads and gRPC for inter-peer communication. Then we compare Rust and Go as the two languages with manual memory management and GC, respectively, which use user-mode threads and TCP for inter-peer communication.

\subsection{Methodology.} We focus our evaluation on two typical workloads: update-heavy workload A (50\% reads, 50\% writes) and read-heavy workload B (95\% reads, 5\% writes) of YCSB~\cite{ycsb-workloads}. To better understand the cost of GC, we run experiments in different configurations that reduce the memory size or the vCPU count. Specifically, on each m5.2xlarge instance running a Replicant leader or follower, we reduce the memory size from 32\,GiB to 8\,GiB, 6\,GiB, and 5\,GiB without changing the vCPU count (8) and run each workload in each of these configurations (we stop at 5\,GiB because our raw dataset size is $\approx$\,4\,GiB). We then reduce the vCPU count from 8 to 4 and 2 without changing the memory size (32\,GiB) and run each workload in each of these configurations. We reduce the memory size for C++ and Rust with \texttt{systemd}'s \texttt{MemoryMax} directive, for Go with the \texttt{GOMEMLIMIT} environment variable, and for Java with the \texttt{-Xmx} parameter of JVM. We reduce the vCPU count by turning vCPUs off via \texttt{sysfs}~\cite{sysfs-core}.

The industry practice for determining the throughput for a server software given a hardware budget is to fix a target tail latency and empirically find the maximum achievable throughput that hits the target tail latency~\cite[p.~4]{dicksites}. We select tail latency as our service-level objective because GC has the most significant impact on tail latencies. Specifically, we set the target 99th percentile latency at 20\,ms, meaning that 99\% of requests are completed within 20\,ms. The value of 20\,ms is 10 times of average latency observed in \autoref{fig:grpc-and-tcp-latency}, where the GC pressure is minimal.

\begin{figure}
    \centering
    \includegraphics[width=\linewidth]{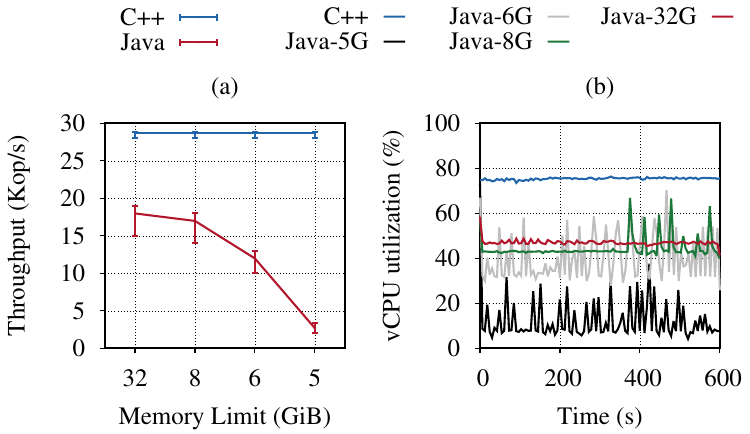}
    \caption{(a) Maximum achievable throughput of C++ and Java with 99th percentile latency of 20\,ms under workload A with 8 vCPUs. C++ remains constant at around 28.7\,Kops/s while Java drops from 18\,Kops/s to 3\,Kops/s when the memory decreases from 32\,GiB to 5\,GiB. (b) The aggregate vCPU utilization of C++ and Java on a Replicant leader during the same experiments; the vCPU utilization of C++ stays the same across experiments. (Java-5G corresponds to vCPU utilization when using 5\,GiB memory size).}
    \label{fig:cpp-java-mem-a}
\end{figure}

We follow the same practice for each of the configurations specified above. In each configuration, we run multiple experiments (each experiment runs for 10 minutes after a 20-second warm-up), adjusting the throughput until we hit our target tail latency. Specifically, we begin with a large throughput setting, and if it exceeds the target tail latency, we repeat the experiment with a lower throughput. For these experiments, we keep the number of client threads fixed at 64 to generate the load and vary the throughput by changing YCSB's target throughput parameter. 

\subsection{Comparing C++ and Java Under Workload A}
\label{sec:cpp-java-a}

\autoref{fig:cpp-java-mem-a}\,{(a)} shows the maximum achievable throughput with the 99th percentile latency of 20\,ms for C++ and Java using different memory configurations and 8 vCPUs. \autoref{fig:cpp-java-mem-a}\,{(b)} shows Replicant leader's aggregate vCPU utilization (i.e., 100\% utilization corresponds to all 8 vCPUs fully utilized) for each experiment. Since C++'s vCPU utilization stays the same (75\%) during all experiments, we show it once.

We first focus on the 32\,GiB memory experiment: \autoref{fig:cpp-java-mem-a}\,{(a)} shows that the throughputs of C++ and Java are 28.7\,Kops/s and 18\,Kops/s, respectively, and \autoref{fig:cpp-java-mem-a}\,{(b)} shows that the vCPU utilizations of C++ and Java are 75\% and 48\%, respectively. This data shows that even when there is ample memory, obviating large GC pauses, Java's throughput is 37\% less than C++'s if a certain tail latency target is required; this happens because some small GC occurs continuously, busying the CPU. Hence, the 38\% lower throughput of Java is a proxy for the CPU cost of GC when there is ample memory---to achieve the same throughput as C++ with the same target tail latency, Java would need more CPU. In other words, the 48\% vCPU utilization in \autoref{fig:cpp-java-mem-a}\,{(b)} is the maximum utilization that will allow Java to reach the target tail latency; if we push the utilization higher by increasing the load, then the tail latency will suffer.

\begin{figure}
    \centering
    \includegraphics[width=\linewidth]{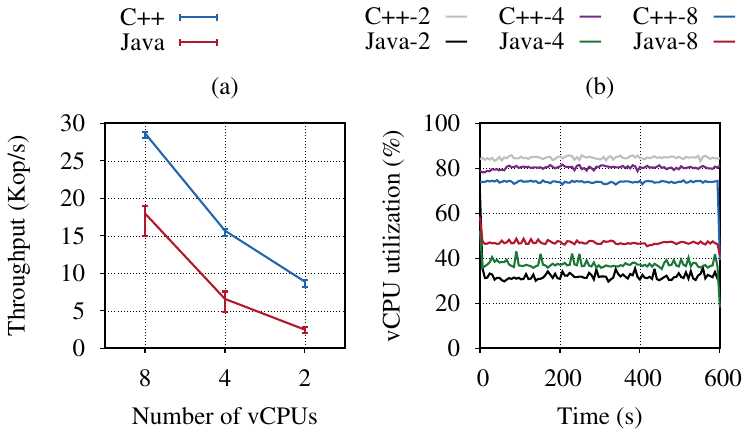}
    \caption{(a) Maximum achievable throughput of C++ and Java with 99th percentile latency of 20\,ms under workload A with different vCPU configurations and 32\,GiB of memory. (b) The aggregate vCPU utilization of C++ and Java on a Replicant leader during the same experiments (Java-2 corresponds to vCPU utilization when using 2 vCPUs).}
    \label{fig:cpp-java-cpu-a}
\end{figure}

Another way to demonstrate this is to reduce the available vCPUs and observe the maximum achievable throughput of Java with the same target tail latency. For example, looking at the 48\% vCPU utilization in \autoref{fig:cpp-java-mem-a}\,{(b)}, one may erroneously assume that halving the vCPU count won't impact the throughput. \autoref{fig:cpp-java-cpu-a} shows the results of vCPU reduction experiments and proves the assumption wrong. \autoref{fig:cpp-java-cpu-a}\,{(a)} shows the maximum achievable throughput with the 99th percentile latency of 20\,ms for C++ and Java using different vCPU configurations and 32\,GiB of memory, and \autoref{fig:cpp-java-cpu-a}\,{(b)} shows the vCPU utilization at a Replicant leader during the same experiments. We observe that when dropping the vCPU count from 8 to 4 while still keeping ample memory (32\,GiB), the throughput of Java drops from 18\,Kops/s to 7\,Kops/s and its utilization drops from 48\% to 38\%, showing that Java needs to keep a large portion of idle CPU around to handle continuous GC while satisfying the target tail latency.

Furthermore, we find that Java is less cost-efficient than C++. Although Java uses less vCPU utilization than C++, cloud service providers charge based on whole virtual machine sizes instead of actual runtime resource utilization, even when all resources are not fully utilized. \autoref{fig:cpp-java-cpu-a} shows that to target the same maximum achievable throughput as C++, Java needs to double the number of vCPUs. For instance, Java using 8 vCPUs reaches a throughput similar to C++ which uses only 4 vCPUs. We further confirm our observation by upgrading the AWS instance from m5.2xlarge (8 vCPUs) to m5.4xlarge (16 vCPUs), where Java achieves 29 Kops/s. However, doubling resource sizes also means doubling costs. Consequently, to achieve the same performance, Java needs to reserve larger machines, leading to higher costs than C++.

Going back to the memory reduction experiments, \autoref{fig:cpp-java-mem-a}\,{(a)} shows that as we keep reducing the available memory size, Java's throughput suffers even more, reaching 3\,Kops/s with 5\,GiB of memory, while C++'s throughput stays the same at 28.7\,Kops/s. This is because, as \autoref{fig:cpp-java-mem-a}\,{(b)} shows, with decreasing memory size, the CPU cost of GC increases further as the garbage collector starts to run more often and for longer periods. Specifically, we see that as the memory size decreases, the GC spikes become more frequent and higher. More importantly, the utilization drops at a higher rate because even more idle CPU is needed to hit the target tail latency. For example, while the utilization drops by only 4\% (43\% $\rightarrow$ 39\%) when going from 8\,GiB to 6\,GiB, it drops by 27\% when going from 6\,GiB to 5\,GiB (39\% $\rightarrow$ 12\%).

Finally, \autoref{fig:cpp-java-cpu-a}\,{(b)} shows that with ample memory, reducing available CPU results in smaller reduction in utilization in Java because less idle CPU is needed to handle infrequent and short GC bursts and more CPU can be used for actual work.

For both experiment types---memory and vCPU reduction---we observe similar trends in the vCPU utilization on Replicant followers, but at a smaller scale; e.g., for memory reduction experiments, C++'s vCPU utilization remains at 30\% while Java's vCPU utilization starts at 12\% and drops to 2\%. We omit the graphs to save space.

\textit{In summary, for an update-heavy workload, when fixing the 99th percentile latency at 20\,ms, C++ achieves 1.7$\times$ higher throughput than Java (28.7\,Kops/s vs. 18\,Kops/s) with ample memory (memory (32\,GiB) to data (4\,GiB) ratio = 8), and 9.6$\times$ higher throughput than Java (28.7\,Kops/s vs. 3\,Kops) with limited memory (memory (8\,GiB) to data (4\,GiB) ratio = 1.25), with the same number vCPUs.}

\subsection{Comparing C++ and Java Under Workload B}
\label{sec:cpp-java-b}
\begin{figure}
    \centering
    \includegraphics[width=\linewidth]{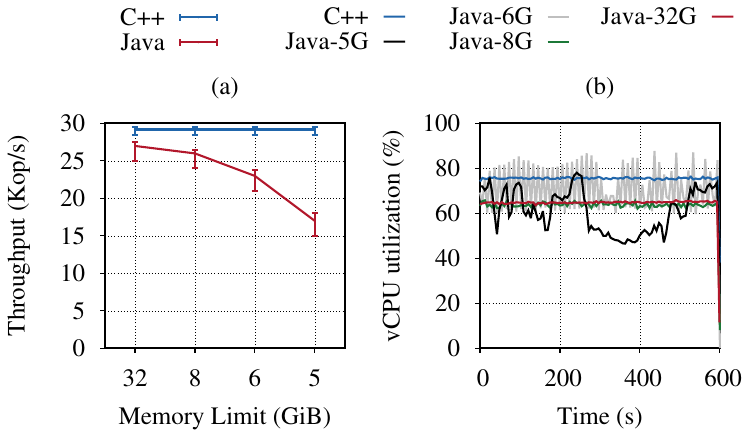}
    \caption{(a) Maximum achievable throughput of C++ and Java with 99th percentile latency of 20\,ms under workload B with different memory configurations and 8 vCPUs. (b) The aggregate vCPU utilization of C++ and Java on a Replicant leader during the same experiments; the vCPU utilization of C++ stays the same across experiments.}
    \label{fig:cpp-java-mem-b}
\end{figure}

\autoref{fig:cpp-java-mem-b} shows the results of memory reduction experiments for workload B. The experiments and the subgraphs are identical to those of \autoref{fig:cpp-java-mem-a} with the only change being switching from update-heavy workload A to read-heavy workload B.

\autoref{fig:cpp-java-mem-b}\,{(a)} shows that in a read-heavy workload, with ample memory the throughputs of C++ and Java are close at 29.2\,Kops/s and 27\,Kops/s, respectively. This is not surprising, because fewer updates translate to less garbage to reclaim for the garbage collector, and more available CPU for the actual work. \autoref{fig:cpp-java-mem-b}\,{(b)} confirms this: with ample memory Java utilizes 60\% of vCPU and thus needs less idle CPU budget to hit the target tail latency, compared to the update-heavy workload A, where the utilization for the same experiment is at 48\%, as \autoref{fig:cpp-java-mem-a}\,{(b)} shows.

\autoref{fig:cpp-java-mem-b}\,{(a)} shows that as we keep reducing the available memory size, Java's throughput starts to suffer---similar to the update-heavy workload (\autoref{fig:cpp-java-mem-a}\,{(a)}) but at a slower rate, reaching 17\,Kops/s with 5\,GiB of memory, while C++'s throughput stays the same at 29.2\,Kops/s. \autoref{fig:cpp-java-mem-b}\,{(b)} shows that unlike update-heavy workload, the vCPU utilization doesn't drop as quickly, because the small amounts of GC due to fewer updates can be handled without requiring a lot of idle CPU to hit the target tail latency. Specifically, when going from 32\,GiB to 8\,GiB, there is a negligible drop in vCPU utilization, and when going from 8\,GiB to 6\,GiB, the aggregate vCPU utilization stays the same although GC spikes are more frequent. It is only when memory becomes extremely limited, at 5\,GiB, we observe a noticeable drop in the vCPU utilization.

For the vCPU reduction experiments of workload B we observe a similar trend to the vCPU reduction experiments of workload A, but with a smaller delta between the throughputs of C++ and Javas at each datapoint. Specifically, while the throughput difference in \autoref{fig:cpp-java-cpu-a}\,{(a)} between C++ and Java averages at 8{,}700\,Kops/s, the throughput difference for workload B averages at 2{,}500\,Kops/s. We omit the vCPU reduction experiment graph for workload B to save space. Similar to workload A experiments, we observe identical trends for both experiment types on Replicant followers, but at a smaller scale.

\textit{In summary, for a read-heavy workload, when fixing the 99th percentile latency at 20\,ms, C++ achieves 1.08$\times$ higher throughput than Java (29.2\,Kops/s vs 27\,Kops/s) with ample memory (memory to data ratio = 8), and 1.7$\times$ higher throughput than Java (29.2\,Kops/s vs 17\,Kops/s) with limited memory (memory to data ratio = 1.25), with the same number of vCPUs.}

\subsection{Comparing Rust and Go Under Workload A}
\label{sec:rust-go-a}
\begin{figure}
    \centering
    \includegraphics[width=\linewidth]{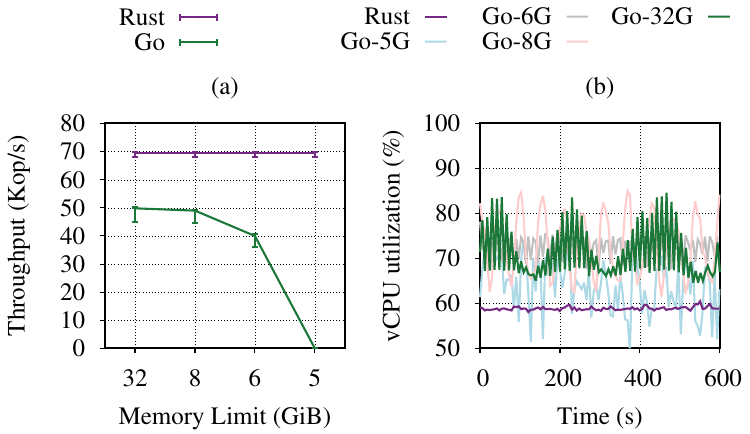}
    \caption{(a) Maximum achievable throughput of Rust and Go with 99th percentile latency of 20\,ms under workload A with different memory configurations and 8 vCPUs. (b) The aggregate vCPU utilization of Rust and Go on a Replicant leader during the same experiments;  the vCPU utilization of Rust stays the same across experiments.}
    \label{fig:rust-go-mem-a}
\end{figure}

\begin{figure}
    \centering
    \includegraphics[width=\linewidth]{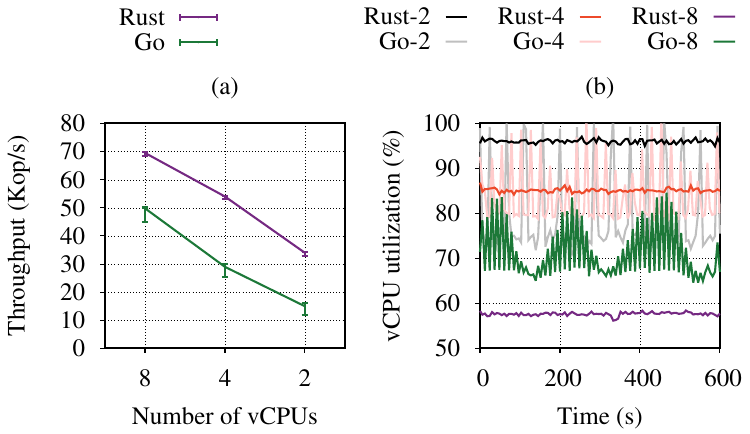}
    \caption{(a) Maximum achievable throughput of Rust and Go with 99th percentile latency of 20\,ms under workload A with different vCPU configurations and 32\,GiB of memory. (b) The aggregate vCPU utilization of Rust and Go on a Replicant leader during the same experiments (Go-2 corresponds to vCPU utilization when using 2 vCPUs).}
    \label{fig:rust-go-cpu-a}

\end{figure}

\autoref{fig:rust-go-mem-a} and \autoref{fig:rust-go-cpu-a} show the results of memory reduction and vCPU reduction experiments, respectively, for Rust and Go under workload A. The experiments and subgraphs are identical to those of \autoref{fig:cpp-java-mem-a} and \autoref{fig:cpp-java-cpu-a}, respectively, with one change: we switch from comparing C++ and Java to comparing Rust and Go.

Focusing on the memory reduction experiments first, \autoref{fig:rust-go-mem-a}\,{(a)} shows the throughput of Rust and Go at 69.5\,Kops/s and 49.8\,Kops/s, respectively, and \autoref{fig:rust-go-mem-a}\,{(b)} shows that the vCPU utilization of Rust and Go is at 58\% and 71\% (when smoothed), respectively. The first observation from these results is that even where there is ample memory, obviating the need for major GC work, Go's throughput is 28\% lower than Rust's if certain tail latency is required; this is because even with ample memory Go's garbage collector works continuously to keep the heap memory overhead and CPU overhead in balance~\cite{go-gc}, as shown by the regular spikes in \autoref{fig:rust-go-mem-a}\,{(b)}.

\autoref{fig:rust-go-mem-a}\,{(a)} shows that as we keep reducing the available memory size, Go's throughput suffers even more while Rust's throughput does not change. To our surprise, when we reduce the memory size to 5\,GiB, Go cannot hit the target 99th percentile latency of 20\,ms even with an abysmally low load of 100 ops/s. Specifically, the 99th percentile latency of Go with a load of 100 ops/s is 154\,ms. This is because with limited memory, Go's garbage collector dominates the CPU leaving little room for actual work.

Moving on to vCPU reduction experiments in \autoref{fig:rust-go-cpu-a}\,{(a)}, we observe that as we decrease the vCPU count from 8 to 4 to 2, the throughput of both Rust and Go drop similarly. In other words, when there is ample memory, the cost of GC remains roughly constant. \autoref{fig:rust-go-cpu-a}\,{(b)} also shows that the vCPU utilization of Rust jumps from 58\% to 85\% to 96\%, as we decrease vCPUs from 8 to 4 to 2. In other words, while Rust has higher throughput than Go, its CPU efficiency decreases with an increasing number of vCPUs. This is due to Log's lock (\autoref{sec:replicant-design}) becoming contended (which is expected since MultiPaxos is not embarrassingly parallel) and Tokio's I/O driver shared among worker threads becoming contended with more vCPUs.%~\cite{noah-tokio}.

\begin{figure}
    \centering
    \includegraphics[width=\linewidth]{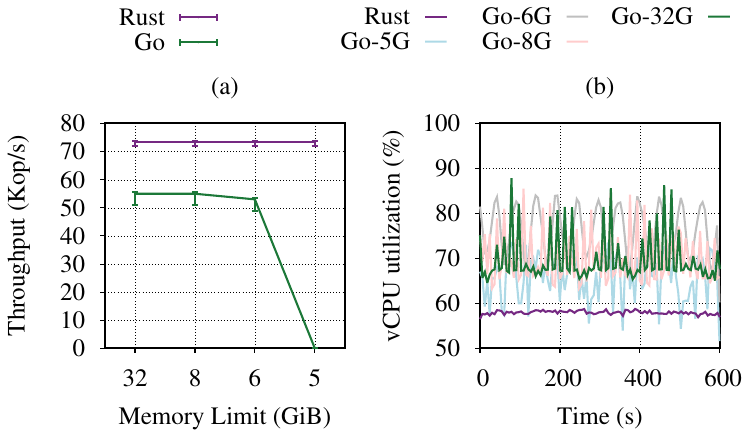}
    \caption{(a) Maximum achievable throughput of Rust and Go with 99th percentile latency of 20\,ms under workload B with different memory configurations and 8 vCPUs. (b) The aggregate vCPU utilization of Rust and Go on a Replicant leader during the same experiments; the vCPU utilization of Rust stays the same across experiments.}
    \label{fig:rust-go-mem-b}
\end{figure}

Moreover, Go is not only less cost-efficiency compared to Rust, but also also less resource-efficient than Rust. To quantify the resource efficiency, we normalize the throughput by runtime vCPU utilization, i.e. throughput-per-unit-of-vCPU-utilization~\cite{scalable-but-wasteful}. Rust achieves a consistent 120 Kops/s/vCPU\% with 8 vCPUs regardless of various memory sizes, while Go experiences a decline from 70 Kops/s/vCPU\% to 55 Kops/s/vCPU\% as the memory size decreases. Even in the vCPU reduction experiment, despite the contention issue in Rust, Rust still maintains higher normalized throughput than Go with the same number of vCPUs.

\textit{In summary, for an update-heavy workload, when fixing the 99th percentile latency at 20\,ms, Rust achieves 1.4$\times$ higher throughput than Go (69.5\,Kops/s vs. 49.8\,Kops/s) with ample memory (memory to data ratio = 8) with the same number of vCPUs; with limited memory (memory to data ratio = 1.25) and with the same number of vCPUs, Go is unable to achieve the target tail latency.}

\begin{figure}
    \centering
    \includegraphics[width=\linewidth]{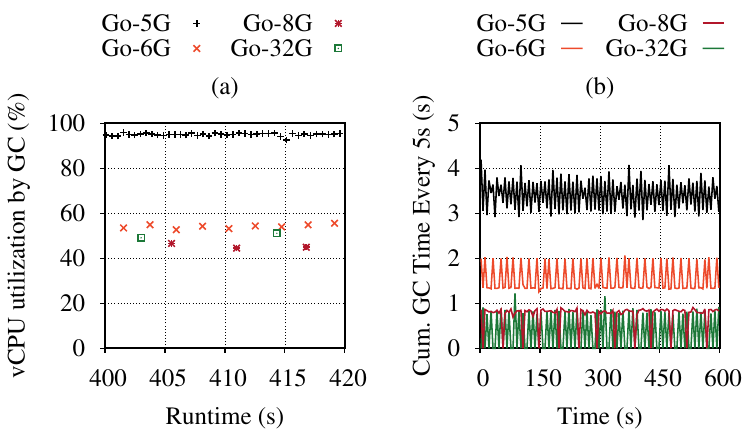}
    \caption{(a) Real-time vCPU utilization of the Go garbage collector under workload A with 8 vCPUs and varied memory configurations. Each data point represented vCPU utilization at a specific time during runtime, with selected runtime ranging from 400 to 420 seconds. (b) The cumulative runtime of Go's GC activities using a 5-second window. This data reflects the aggregation of the duration of all GC activites happened within each 5-seond interval. (Go-5G corresponds to the 5\,GiB memory size).}
    \label{fig:go-gc-trace}
\end{figure}

\subsection{Comparing Rust and Go Under Workload B}

\autoref{fig:rust-go-mem-b} shows the results of memory reduction experiments for workload B. The experiments and the subgraphs are identical to those of \autoref{fig:rust-go-mem-a}, with the only change being switching from update-heavy workload A to read-heavy workload B.

\autoref{fig:rust-go-mem-b}\,{(a)} shows that in a read-heavy workload, even with ample memory, there is a significant difference between Rust's and Go's throughputs at 73.5\,Kops/s and 55\,Kops/s, respectively. We are careful not to compare this result to C++ vs. Java result for the same workload (\autoref{sec:cpp-java-b}) and draw conclusions about the quality of garbage collectors: Java uses OS threads with gRPC and reaches roughly half the throughput of Go, resulting in half the updates and half the garbage to reclaim.

\autoref{fig:rust-go-mem-b}\,{(a)} shows that as we keep reducing the available memory size, Go's throughput stays relatively stable until we reach the 5\,GiB. Surprisingly, even for a read-heavy workload, Go cannot hit the target 99th percentile latency of 20\,ms with an abysmally low load of 100 ops/s. Specifically, the 99th percentile latency of Go with a load of 100 ops/s is 156\,ms.

For the vCPU reduction experiments of workload B, we observe a similar trend to the vCPU reduction experiments of workload A, with an almost identical delta of 21\,Kops/s between the throughputs of Rust and Go. We omit the vCPU reduction experiment graph for workload B to save space.

\textit{In summary, for a read-heavy workload, when fixing the 99th percentile latency at 20\,ms, Rust achieves 1.33$\times$ higher throughput than Go (73.5\,Kops/s vs. 55\,Kops/s) with ample memory (memory to data ratio = 8) with the same number of vCPUs; with limited memory (memory to data ratio = 1.25) and with the same number of vCPUs, Go is unable to achieve the target tail latency.}

\subsection{Confirming the GC Overhead}

\begin{figure}
    \centering
    \includegraphics[width=\linewidth]{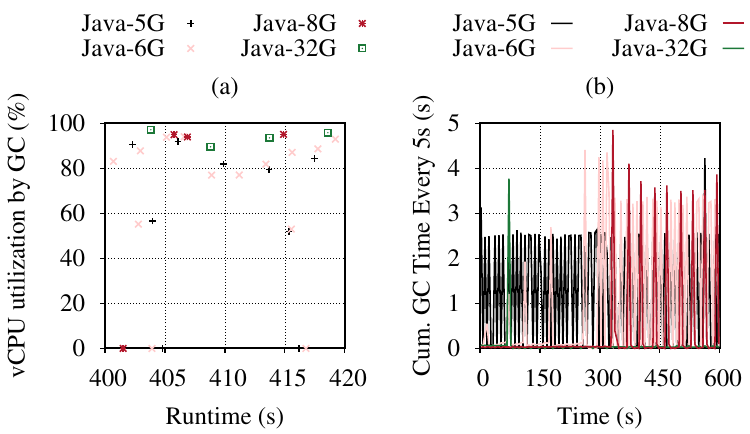}
    \caption{(a) Real-time vCPU utilization of the Java garbage collector under workload A with 8 vCPUs and varied memory configurations, with selected runtime ranging from 400 to 420 seconds. (b) The cumulative runtime of Go's GC activities within each 5-seond interval. (Java-5G corresponds to the 5\,GiB memory size).}
    \label{fig:java-gc-trace}
\end{figure}

How do we know that the difference in resource usage in these implementations observed above is actually due to GC? To answer this question, we enable the \texttt{GODEBUG=gctrace} option for Go and \texttt{-Xlog:gc*} for Java, allowing us to capture detailed GC activities during runtime. Specifically, once we determine the maximum achievable throughput under workload A and different memory configurations for Go and Java, we repeat the experiment with the GC trace option enabled, maintaining identical experiment settings.

\autoref{fig:go-gc-trace}\,{(a)} shows the \textbf{instantaneous} vCPU utilization by GC and the frequency of GC events at a Replicant leader in Go. Each point represents a GC event occurrence during the runtime, with denser points indicating more frequent GC activities. Given the high frequency and even distribution of these events, we select a segment of runtime data to present the graph more effectively. Note that, in contrast to previous vCPU utilization graphs, \autoref{fig:go-gc-trace}\,{(a)} exclusively shows the \textbf{instantaneous} utilization caused by GC, capturing the vCPU usage during GC activities, rather than throughout the entire runtime. The data reveals that the \textbf{instantaneous} vCPU utilization caused by GC is significantly higher than the rest with 5\,GiB of memory (95\%), followed by the 6\,GiB of memory limitation (55\%).

When we reduce the memory size from 6\,GiB to 5\,GiB, the frequency of GC activities grows 5 times higher; this is attributed to the necessity for more frequent GC clean-up events to free up spaces for new memory allocations. Considering that the individual GC durations vary, we calculate the total duration of all GC activities within a 5-second interval, which is shown in \autoref{fig:go-gc-trace}\,{(b)}. Our key observation is that the garbage collector runs along with the application for over 60\% of the runtime with 5\,GiB of memory and almost 40\% of the runtime with 6\,GiB of memory. This explains why Go cannot hit the target 99th percentile latency under 5\,GiB, as the majority of the available vCPUs are occupied by GC activities. 

Despite similar vCPU utilization between 8\,GiB and 32\,GiB of memory, the frequency of GC activities doubles when we decrease memory size from 32\,GiB to 8\,GiB, which is consistent with the total GC duration under a 5-second interval shown in \autoref{fig:go-gc-trace}\,{(b)}. Furthermore, even with ample memory, Go still has nearly 50\% \textbf{instantaneous} vCPU usage for GC, highlighting that its effect becomes more significant as the memory size reduces.

\autoref{fig:java-gc-trace} shows the \textbf{instantaneous} vCPU utilization and cumulative duration of all GC events every 5 seconds for Java. Unlike Go, Java’s GC traces do not offer comprehensive data on CPU usage. Consequently, \autoref{fig:java-gc-trace}\,{(a)} shows the vCPU utilization of the clean-up phase only, excluding the concurrent mark phase in Java GC. Although Java spends less time on GC compared to Go, it still exhibits notable vCPU spikes regardless of memory constraints, as illustrated in \autoref{fig:java-gc-trace}\,{(a)}. 

It is important to note that we are not conducting comparison or analysis on each garbage collector. Java’s default G1 garbage collector aims to balance between minimizing pause time and higher throughput~\cite{g1gc}. It is consistent with our key observation in \autoref{sec:cpp-java-a}. The \textbf{instantaneous} vCPU utilization is normalized in \autoref{fig:cpp-java-mem-a}\,{(b)} because the GC duration is short. However, \autoref{fig:java-gc-trace}\,{(a)} shows that Java needs to keep available CPU to deal with those regular high GC spikes. Furthermore, the cost of full GC remains huge. When the full GC is triggered, it results in more time spent on clean-up and marking in the latter runtime shown in \autoref{fig:java-gc-trace}\,{(b)}, leading to much longer 99th percentile latency. These longer marking phases and full GC events inevitably cause significantly higher tail latency in Java.

\subsection{Virtualization Overhead}
\label{sec:virt-overhead}

We conclude this section by exploring how the virtualization overhead on the cloud affects languages. To this end, we repeat the workload A experiments (\autoref{sec:cpp-java-a}, \autoref{sec:rust-go-a}) with no resource limitations on m5.2xlarge instances and on m5.metal instances (96 CPUs, 384\,GiB memory) with 8 online CPUs (we turn off the rest via \texttt{sysfs}~\cite{sysfs-core}) and with memory size reduced to 32\,GiB---to match the m5.2xlarge configuration. Using \texttt{lscpu}, we confirm that both instance types have identical CPUs (Intel Xeon Platinum 8259CL CPU at 2.50GHz).  

\autoref{tbl:virt-overhead} shows the maximum achievable throughputs with 99th percentile latency of 20\,ms on bare metal and virtualized instances and the percentage decrease in throughputs. We observe that C++ suffers the most from the virtualization overhead followed by Java, with drops of 20\% and 10\% in throughputs, respectively. Go and Rust, on the other hand, have a negligible drop in throughput. While it appears that the languages with user-mode threads are at an advantage, we leave a data-backed explanation of this phenomenon as a future work. Our key observation from these results is that language comparisons on bare metal may not translate to the cloud.

\begin{table}[t]
 \centering {
  \setlength{\tabcolsep}{2.3pt}
  \begin{tabular*}{\columnwidth}{rrrrr} \toprule
    & C++          & Java & Go & Rust \\\midrule
   m5.metal  & 36\,Kops/s         & 20\,Kops/s & 50\,Kops/s & 70\,Kops/s \\\
   m5.2xlarge  & 28.7\,Kops/s         & 18\,Kops/s & 49.8\,Kops/s & 69.5\,Kops/s \\\midrule
   Decrease  & 20\%         & 10\%  & 0.004\% & 0.007\% \\\
  \end{tabular*}
 }
 \caption{
   Maximum achievable throughput of Replicant implementations with 99th percentile latency of 20\,ms under workload A on m5.metal instances with 8 online CPUs and memory limited to 32\,GiB and on m5.2xlarge instances, including the percentage decrease in throughput.
}
 \label{tbl:virt-overhead}
\end{table}

\section{Language Experience and Discussion}
\label{sec:discussion}
In our subjective experience of building Replicant in multiple languages, we avoided language-specific optimizations and tried to use as many out-of-the-box features of languages as possible. While each language can be tuned with expert knowledge of compiler or garbage collector internals, it would require a level of sophistication that is not available to most programmers. To put our experience into context, the programming experience of our team members includes 10+ years of C++, 7+ years of Java, and 4+ years of Go---some of these at hyperscalers and medium-sized software companies. We had no prior Rust experience, but we studied it~\cite{rust-book, programming-rust} for a month before starting our Rust implementation.

Go, by far, was the quickest language to build Replicant in, with the least friction and without resorting to third-party libraries, and it provided a decent performance from the get-go. Java and C++ were next in terms of ease of development and project completion time. For C++, we additionally integrated AddressSanitizer~\cite{asan} to our development workflow, which found a few a use-after-free bugs at different stages of the development. Rust was the most challenging language to work with initially, and not only because we lacked prior experience, but because it requires doing routine things differently; e.g., in most languages, knowing how to use a mutex and a structure is enough for implementing a monitor~\cite{monitor}; in Rust, on the other hand, it requires an unobvious pattern~\cite{shared-mutable-state}.

To overcome our initial friction with Rust, we purchased a one-month Rust consultancy (for 512 \texteuro s), which helped us to overcome early roadblocks in our implementation; we got introduced to advanced techniques and idioms not covered in the books, which were essential to our implementation. As we became comfortable with Rust, it started to have a significant impact---in a good way---on our project. It forced us to rethink our interfaces (and thereby our design) to satisfy the complaints of the Rust compiler, which made the interfaces simpler, less sloppy, and less error-prone, and we had to go back and implement those changes in other languages. Finally, profiling efforts for identifying the H2 and \texttt{ptmalloc} bottlenecks (\autoref{sec:tcp-fork}) were substantial.

Focusing on the manual memory management overhead of the languages, while C++ successfully hides most of this overhead from programmers using the RAII idiom~\cite{raii}---our C++ codebase never explicitly allocates or frees memory---we still had to think about the lifetime and ownership of objects during the design. Furthermore, since there is no memory-safety guarantee, we had to integrate AddressSanitizer~\cite{asan} into our development workflow.

Rust, unlike C++, (largely) guarantees memory safety and data-race-freedom at compile time. These guarantees, however, come at the cost of a steep learning curve with unfamiliar and perplexing concepts and a generally slower pace of development because the compiler will not allow ``hacking away'' without thinking through the lifetimes and concurrency implications of the new code---and thinking is hard and takes time. The assistance we got from consultants helped us to overcome the initial impedance with the language, and, e.g., we replaced gRPC in Replicant to use TCP for inter-peer communication (\autoref{sec:tcp-fork}) relatively quickly (it was not trivial and required a redesign). After many thousands of lines of Rust, our subjective opinion is that we feel comfortable building large projects in Rust, but others have reported a different experience~\cite{welsh-rust}.

\begin{figure}
    \centering
    \includegraphics[width=\linewidth]{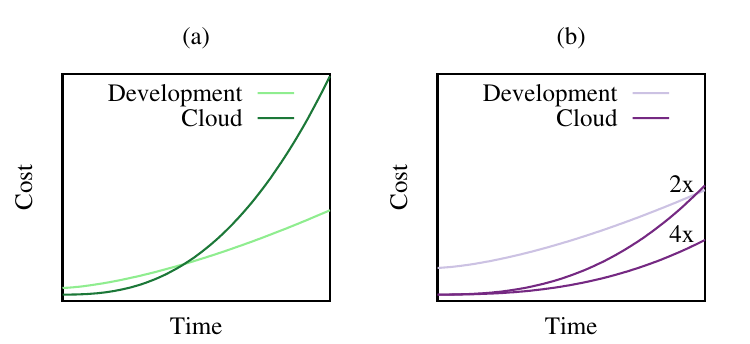}
    \caption{A simplistic model for the long-term development and cloud costs of building a cloud system in a language (a) with GC and (b) with manual memory management.}
    \label{fig:cost-efficiency}
\end{figure}

To sum up, using a language with manual memory management does have a higher development cost than using a language with GC. Therefore, our observation is that the choice of a programming language for building a cloud system is a trade-off between the long-term development cost and cloud cost. We illustrate this trade-off with a simplistic model in \autoref{fig:cost-efficiency}. \autoref{fig:cost-efficiency}\,{(a)} shows the long-term costs of building a cloud system in a language with GC: the development cost starts higher than the cloud cost and grows slowly; the cloud cost starts low, but as the system grows in popularity, the cloud cost ends up dwarfing the development cost in the long run. \autoref{fig:cost-efficiency}\,{(b)} shows the same for a language with manual memory management, where the development cost starts 2$\times$ higher than \autoref{fig:cost-efficiency}\,{(a)} and grows slowly; we show how the cloud cost would look like with the same growth rate as in \autoref{fig:cost-efficiency}\,{(a)} if the system were using 2$\times$ fewer resources and 4$\times$ fewer resources.

In conclusion, our observation is that if a cloud system is expected to grow to a large volume of users, using a language with manual memory management would significantly reduce long-term cloud cost despite high development cost. Conversely, if the system is expected to grow up to a certain small bound, then using a language with GC would reduce long-term development cost.

%Finally, orthogonal to GC, it appears that user-mode threads experience less virtualization overhead and deliver higher throughput, which makes them a better fit over the OS threads for building cloud systems.

\section{Related Work}
Our work is the first study that implements a large cloud system in languages with GC and manual memory management to systematically quantify the overheads of GC.

%\textbf{Language choice.} Previous works discussed the impact of language choice on both using micro and macro benchmarks. Cutler et al.\cite{cutler} compare the kernel implemented in Go and the Linux kernel (implemented in C). They mainly emphasize the security and ease of implementation aspect of Go while touching on the performance cost incurred by the Go garbage collection and safety checks. 
%Lion et al.\cite{lion} claim to be the first work analyzing and comparing the multiple widely used runtimes along with the instrumented benchmarks. They both study overheads and the advantages of having a managed runtime. While the type and bounds checking, interpreter, and GC write barriers incur an overhead, applications benefit from improved cache locality due to the GC and have better scalability(in Go) due to the runtime scheduler. However, their study is limited to the in-house bare metal servers. As we show in this work, it requires significant effort and mastery to tune the implementation in the given language, which is also highlighted by Hundt\cite{hundt}. Similar to our work, it implements a selected algorithm in different languages and shows how one can benefit from language-specific optimizations. However, more engineering effort is required to tune the implementation, especially the garbage collection.

\textbf{Language choice.} Several prior works use micro and macro benchmarks to discuss and showcase the impact of language choice on systems. Cutler et al.~\cite{cutler} compare the kernel implemented in Go and the Linux kernel (implemented in C). The authors emphasize Go's security and ease of implementation while briefly touching on the performance cost incurred by the Go garbage collection and safety checks. Lion et al.~\cite{lion} present the first study analyzing and comparing multiple widely used languages with the instrumented benchmarks. In this work, the authors introduce LangBench, a new benchmark suite for evaluating language performance, and use the suite to study the overheads and the advantages of having a managed runtime. Like us, the LangBench uses idiomatic and identical implementations across languages. However, all their tested problems are single-machine applications and may not represent all workloads in the modern cloud. The study finds that while the type and bounds checking, interpreter, and GC write barriers incur an overhead, applications benefit from improved cache locality due to GC and have better scalability (in Go) due to the runtime scheduler. The LangBench results relied on the in-house bare metal servers, and as we show in our work, the virtualization used in the cloud can significantly impact performance. Similar to other studies, the experience report by Hundt~\cite{hundt} compares identical micro-benchmark implementations in various languages. The report, however, highlights the dramatic impact of language-specific optimizations and tuning when deviating from generic implementation. Such potential for different performance mirrors our experience with the first gRPC-based implementation of Replicant. 

%\textbf{Efficiency in the cloud.} Energy efficiency in software has gained traction in the past few years. Most works\cite{pereira, couto,PEREIRA2021102609} have analyzed various languages for their energy consumption using different benchmark problems\cite{computer-language-benchmark-game, rosetta-code}. Maas et al.\cite{maas} advocate for rethinking language runtimes to have better performance in modern data centers. They argue against the use of JIT compilation in the cloud due to the short-living nature of cloud workloads. While they also highlight the performance overhead of GC, they don't favor the idea of improving or replacing the GC. Instead, they propose to have lightweight abstractions to reduce the number of objects. Anderson et al. \cite{anderson2022treehouse}draw attention to the energy efficiency in the cloud by proposing the Treehouse project. Their main goal is to have a new software infrastructure treating energy and carbon as first-class resources together with traditional computing resources. In a recent report by Amazon\cite{aws-rust}, the need for energy-efficient implementations is emphasized for sustainable computing. It advocates that one can have a C-like performance by using Rust together with its sustainability and security benefits. 

\textbf{Efficiency in the cloud.} The efficiency of languages and runtimes has gained traction in the past few years. Most works~\cite{pereira, couto,PEREIRA2021102609} have analyzed various languages for their energy consumption using different benchmark problems\cite{computer-language-benchmark-game,rosetta-code}. Maas et al.~\cite{maas} advocate rethinking language runtimes for better performance in modern data centers. The authors argue against using JIT compilation in the cloud due to the short-lived nature of cloud workloads. While they also highlight the performance overhead of GC, they don't favor the idea of improving or replacing the GC. Instead, they propose to have lightweight abstractions to reduce the number of objects. The Treehouse project~\cite{anderson2022treehouse} calls for a new software infrastructure that treats energy and carbon as first-class resources along with traditional computing resources. A recent report by Amazon\cite{aws-rust} emphasizes the need for energy-efficient implementations for sustainable computing. The report advocates that one can have a C-like performance with Rust while getting more sustainability and security benefits. Matte et al.~\cite{scalable-but-wasteful} investigates the resource efficiency of popular SMR protocols in the cloud. The authors conclude that fast systems may not always be the most efficient, as speed can be achieved by the ability to consume more resources instead of their efficient use. Unlike that study, we compare the efficiency of a specific SMR algorithm implemented in different languages and runtime paradigms.

\section{Conclusion}
In this paper, we quantify the overhead of running a state machine replication system for cloud systems written in a language with GC. To this end, we (1) design from scratch a canonical cloud system---a distributed, consensus-based, linearizable key-values store, (2) implement it in C++, Java, Rust, and Go, (3) evaluate implementations under an update-heavy and read-heavy workloads on AWS under different resource constraints while trying to hit the maximum throughput with a fixed low tail latency. Our results show that even with ample memory, GC has a non-trivial cost, and with limited memory, languages with memory management can achieve an order of magnitude higher throughput than the languages with GC on the same hardware. Our key observation is that if a cloud system is expected to grow to a large volume of users, building the system in a language with manual memory management and thereby paying a higher development cost than using a language with GC may result in a significant cloud cost savings in the long run.

\bibliographystyle{ACM-Reference-Format}
\bibliography{systems}

\end{document}